\documentclass[aps,pra,groupedaddress, twocolumn]{revtex4-1}
\usepackage{hyperref}
\usepackage{amsmath}
\usepackage{amssymb}
\usepackage[pdftex]{graphicx}
\usepackage{microtype}
\usepackage{verbatim}
\usepackage[colorinlistoftodos]{todonotes}
\presetkeys{todonotes}{inline,backgroundcolor=yellow}{}
\usepackage[version=3]{mhchem}

\makeatletter
\DeclareRobustCommand{\element}[1]{\@element#1\@nil}
\def\@element#1#2\@nil{%
  #1%
  \if\relax#2\relax\else\MakeLowercase{#2}\fi}
\makeatother

\newcommand{\mrm}{\mathrm}

\newcommand{\ket}[1]{{\left| {#1} \right\rangle}}

\newcommand{\Bsca}{\mathcal{B}}

\newcommand{\e}{$^{5}D_0$~}
\newcommand{\f}{$^{7}F_1$~}
\newcommand{\g}{$^{7}F_0$~}
\newcommand{\srf}{\ce{SrF_2}\,}

\newcommand{\Euion}{\ce{Eu^{3+}}\,}
\newcommand{\Sm}{\ce{Sm^{2+}}\,}

\newcommand{\Clocktrans}{$^7F_0 - {}^5D_0$}

\usepackage{subcaption} 
\usepackage{cleveref}

\begin{document}
\title{Direct observation of a highly forbidden optical transition in Sm:\srf}
\author{Mohit Verma}
\author{Luca Talamo}
\altaffiliation{Present address: Department of Physics, University of Colorado, Boulder CO, USA.}
\author{Hiromitsu Sawaoka}
\altaffiliation{Present address: Department of Physics, Harvard University, Cambridge MA, USA.}
\author{Amar C. Vutha}
\email{vutha@physics.utoronto.ca}
\affiliation{Department of Physics, University of Toronto, Canada M5S 1A7}

\begin{abstract}
The $4f^6$ \g $\to 4f^6$ \e intra-configuration transition in Sm:\ce{SrF_2} is forbidden for Sm$^{2+}$ ions in the octahedrally symmetric substitution sites in \ce{SrF_2}. We report the direct observation of this transition using laser-induced fluorescence at cryogenic temperatures, and measurements of the excited state lifetime and the excitation cross section. To the best of our knowledge, this optical transition has the longest lived excited state ever observed 
in a solid.
\end{abstract}

\maketitle

\section{Introduction}
Rare-earth ions (REI) doped into solid-state hosts have been studied for many decades due to their unusual optical properties \cite{rubio1991doubly,Wybourne2004,Hull2005}. More recently, REI-doped solids have been used for quantum information processing \cite{simon2010quantum,thiel2011rare,Zhong2015,dibos2018atomic}, and to stabilize narrow-linewidth lasers (e.g., \cite{nilsson2005solid,thorpe2011frequency,Cook2015}). REI-doped solids offer the advantage that very large numbers of atoms, typically $\gtrsim 10^{16}$ in a cm-sized crystal, can be studied without requiring laser-cooling or trapping. In addition, many applications of REI take advantage of the intra-configuration $4f \to 4f$ transitions in these ions, which are shielded by their closed $5s$ and $5p$ shells from interactions with the host lattice: this shielding leads to remarkably narrow optical spectral lines even in the solid-state \cite{equall1994ultraslow,nilsson2005solid,thorpe2011frequency,Cook2015}. Extremely long coherence times for electron and nuclear spins have also been demonstrated in these systems \cite{siyushev2014coherent, Zhong2015}. 

The narrowest observed optical transition in a REI-doped solid thus far is the \Clocktrans\ transition in Eu$^{3+}$ doped into \ce{Y_2SiO_5} (YSO), with homogenous linewidth $\gamma_h = 2 \pi \times$122 Hz and excited state lifetime $\tau \approx 2$ ms at cryogenic temperatures \cite{equall1994ultraslow,konz2003temperature}. The two states involved in this clock transition have no electronic magnetic moments, and low differential sensitivity to crystal field effects. This clock transition has been applied to laser frequency stabilization using spectral hole-burning, reaching fractional frequency stabilities comparable to those achieved with high-finesse optical cavities \cite{leibrandt2013absolute,Cook2015}. However the \Clocktrans\ transition in Eu:YSO is spectrally broad, with an inhomogeneous linewidth $\Gamma_\mrm{inh} \approx 2 \pi \times$1 GHz \cite{Ferrier2016}. Inhomogenous broadening is typically caused by impurities and strains within the crystal that lead to ions at different locations experiencing a range of different local electromagnetic fields \cite{stoneham1969shapes,Cook2015, oswald2018characteristics}. 
Further, the nuclear magnetic dipole and electric quadrupole moments of europium open up pathways for coupling between the \Euion ions and their environment, including adjacent Y$^{3+}$ ions in YSO \cite{equall1994ultraslow}. While such inhomogeneously broadened clock transitions are useful for laser stabilization using spectral hole-burning, they cannot be used as an absolute frequency reference.

In order to explore the fundamental limits to line-broadening in such shielded REI clock transitions, we have been investigating the \Clocktrans\ transition in the isoelectronic ion, \ce{Sm^{2+}}. Samarium has a number of stable isotopes, with mass numbers $A = (144, 147, 148, 149, 150, 152, 154)$ and natural abundances of (3, 15, 11.2, 13.8, 7.4, 26.7, 22.8) \% respectively \cite{Wasserburg1981}. The two odd isotopes ($A = {147,149}$) have nuclear spin $I=7/2$, while the even $A$ isotopes all have zero nuclear spin. The zero spin isotopes are particularly interesting as they could have lower inhomogeneous broadening due to the absence of nuclear moments. When doped into the \ce{SrF_2} host lattice, \Sm (ionic radius $r_\mrm{ion}$ = 141 pm \cite{Jia1991}) substitutes for the similar-sized \ce{Sr^{2+}} ($r_\mrm{ion}$ = 140 pm \cite{Shannon1976}) leading to lower doping-induced strain and potentially better isolation of the clock transition. While this system thus holds promise for realizing an optical frequency reference, the \Clocktrans\ transition in Sm:\ce{SrF_2} had not been studied in sufficient detail prior to this investigation. The \Clocktrans\ transition was only ever observed using weakly allowed decays after indirectly populating the $^5D_0$ state \cite{wood1962absorption,mahbubul1967lifetime}, and so existing estimates for its frequency were not sufficiently precise to allow continuous-wave laser excitation of this transition. The lifetime of the excited state (which sets a lower limit to $\gamma_h$) and the line strength were similarly unknown to high precision. These parameters are necessary to evaluate the feasibility of using Sm:\ce{SrF_2} as an optical atomic clock.

In this work, we report the first direct observation of the \Clocktrans\ transition in Sm:\ce{SrF_2}. By directly exciting the transition, we have characterized the dynamics of laser-induced fluorescence on the clock transition, and measured the excited state lifetime and the excitation cross section. 

\begin{figure*}
{\centering
\includegraphics[width=.95\textwidth, scale = 1.0]{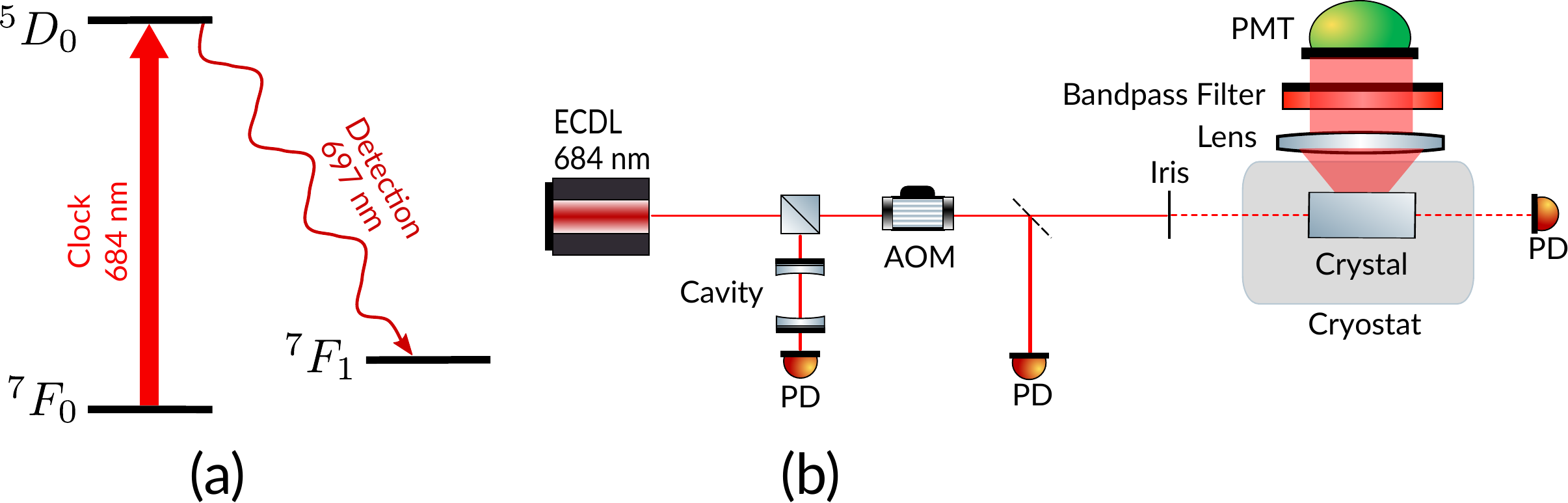}\textit{}
}
\caption[width=\textwidth]{(a) Schematic of the relevant energy levels of Sm$^{2+}$ in Sm:\ce{SrF_2} (see Ref.\ \cite{wood1962absorption} for a more detailed compilation). The \Clocktrans\ clock transition was directly excited at 684 nm, and detected using spontaneously emitted fluorescence from the \e state, which primarily decays to \f by a magnetic dipole ($M1$) transition at 697 nm. All three of the energy levels shown here arise from the $4f^6$ configuration. (b) Schematic of the apparatus used for the measurements reported here. ECDL is an external-cavity diode laser, AOM is an acousto-optic modulator, PMT is a photomultiplier tube, PD are photodiodes.}
\label{fig:LevelStructure}
\end{figure*}

\section{Apparatus}
The measurements were performed on a Sm:\srf crystal (0.1\% nominal Sm concentration, no isotopic enrichment). The electronic level structure of this system is shown schematically in Fig. \ref{fig:LevelStructure}(a). The crystal was secured to the cold plate of a liquid helium cryostat using 
copper clamps, with indium foil used on all the metal-crystal interfaces to ensure good thermal contact. All the measurements reported here were made at 4.2 K, except for measurements of the temperature dependence of the excited state lifetime described below. A schematic of the apparatus is shown in Fig. \ref{fig:LevelStructure}(b).

A 684 nm interference-filter-stabilized external cavity diode laser (ECDL) was used to probe the clock transition. The ECDL was optimized to have a 12 GHz scan range. The intensity of the 684 nm probe laser was controlled using an acousto-optic modulator (AOM), and the fluorescence signal at 697 nm was detected using a photomultipler tube (PMT). Light scattered into the PMT from the probe laser was suppressed using a bandpass filter centered at 697 nm. For the cross section measurements described below, the linewidth of the probe laser was reduced to 1 kHz by locking it to an optical cavity.

\section{Measurements}
\begin{figure}
\centering
    \includegraphics[width=\columnwidth]{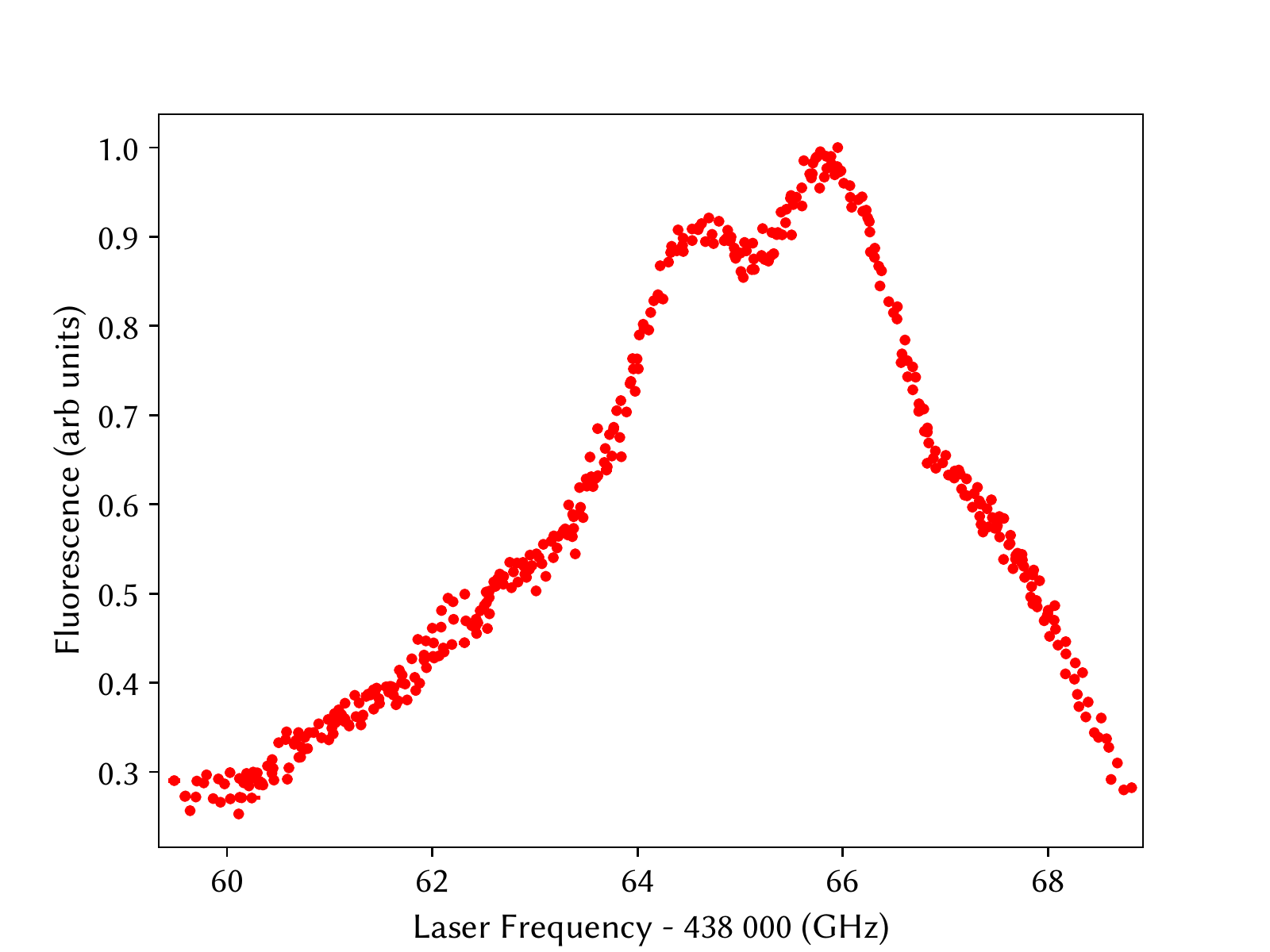}
\caption{The \Clocktrans\ transition in Sm:\ce{SrF_2} observed using laser-induced fluorescence. The vertical axis shows the laser-induced fluorescence signal detected at 697 nm, normalized to the laser power at each frequency. The frequency axis was calibrated using a commercial wavemeter. We attribute the two peaks to inhomogeneously broadened \Clocktrans\ lines from $^{149}$Sm (centered at 438064.3 GHz) and $^{147}$Sm (centered at 438065.6 GHz).}
\label{fig:laser_induced_fluorescence}
\end{figure}

At the outset of this investigation, a number of conflicting values for the frequency of the \Clocktrans\ transition had been reported in the literature. Wood and Kaiser \cite{wood1962absorption} reported the transition wavenumber as 14616 cm$^{-1}$ based on a weak emission line, whereas G\^{a}con \emph{et al.}\ \cite{gacon1993f} excited ions up to the $^5D_0$ state in a two-photon configuration using a pulsed laser and reported 14603 cm$^{-1}$ as the transition wavenumber. Macfarlane and Meltzer \cite{macfarlane1985spectral} directly observed satellite lines at 14620 cm$^{-1}$ from ions located at low-symmetry defect sites, where the transition is more strongly allowed due to admixture of energy levels by the asymmetric crystal field. After initially failing to observe the transition at 14616 cm$^{-1}$ or 14603 cm$^{-1}$, we were able to directly excite the transition in the neighborhood of 14612 cm$^{-1}$ after a thorough search, as shown in Fig.\ \ref{fig:laser_induced_fluorescence}. (To avoid any ambiguities between air versus vacuum wavenumbers, which are likely responsible for the discrepancy with Ref.\ \cite{wood1962absorption}, we only use frequency units below.)

\begin{figure}[h!]
\includegraphics[width = \columnwidth]{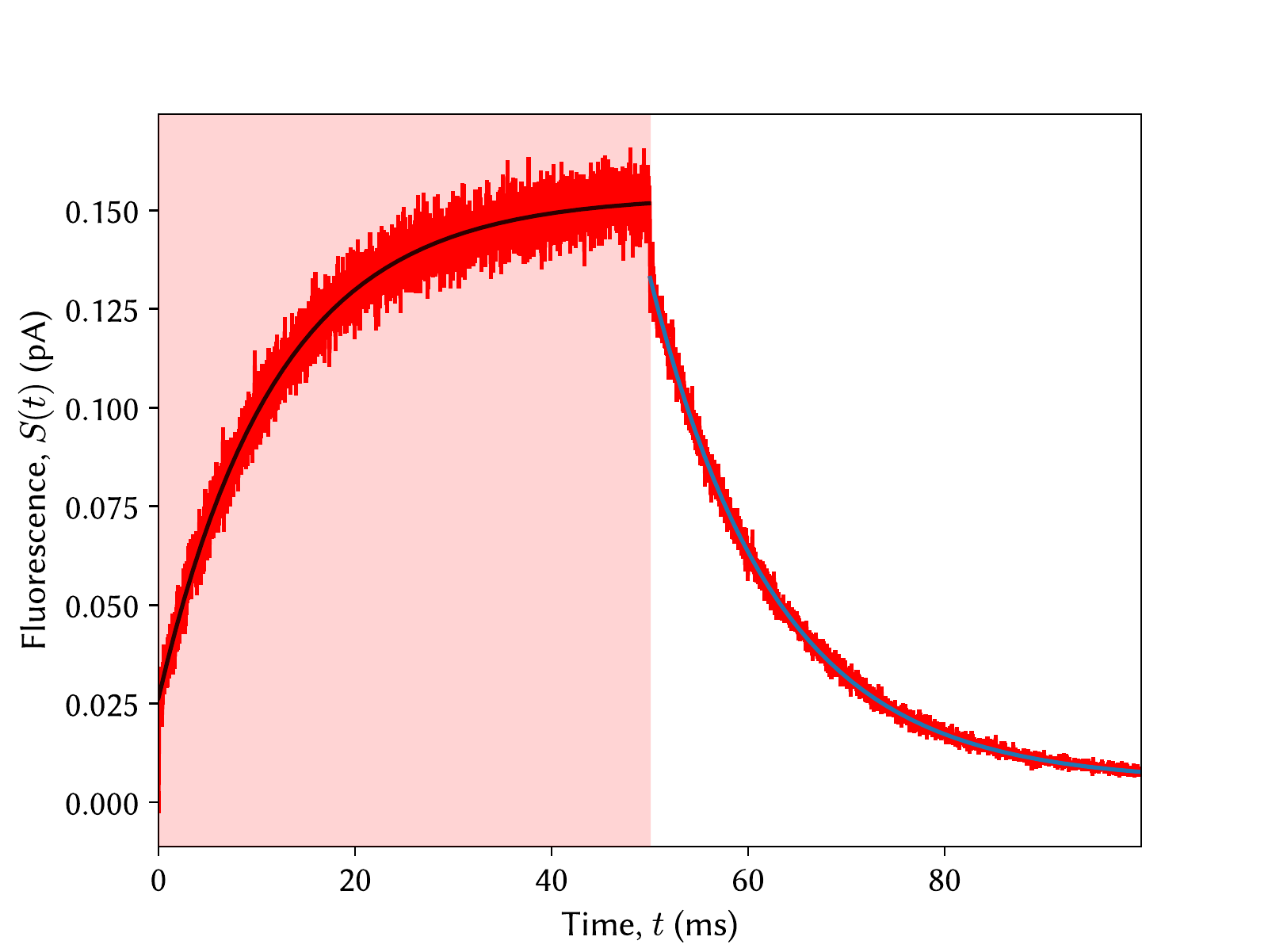}
\centering
\caption{Laser-induced fluorescence measured on the $^5D_0 \to {}^7F_1$ line at 697 nm. The photocathode current is plotted here as a function of time. The probe laser pulse at 684 nm lasts for the first 50 ms (shaded). The solid lines are fits to rising and falling exponentials with the same $1/e$ lifetime $\tau$. The constant offset from residual scattered laser light is removed for clarity. The decay of the fluorescence in the dark (between 50-100 ms) from many such measurements is shown in Fig.\ \ref{fig:RedDecay}.
} 
\label{fig:saturation}
\end{figure}

For the measurement in Fig.\ \ref{fig:laser_induced_fluorescence}, we scanned the frequency of the 684 nm ECDL in 100 MHz steps between 438056 GHz and 438070 GHz. We used 3 mW of laser power and illuminated a 400 $\mu$m spot (1/$e^2$ radius) along the 10 mm length of the crystal. The intensity of the probe laser was constant over the scan to better than 5\%, as measured using monitor photodiodes. At each frequency, 16 laser pulses (typically 50 ms long, repeated once every 100 ms) were sent using the AOM and the 697 nm fluorescence decay signal was measured both during the excitation as well as in the dark. The $^5D_0$ state decays almost exclusively to the $^7F_1$ state, as expected for \Sm ions located in sites with octahedral ($O_h$) symmetry \cite{tanner2013some}. A typical fluorescence time trace is shown in Fig.\ \ref{fig:saturation}. The fluorescence signal shown in Fig.\ \ref{fig:laser_induced_fluorescence} was normalized to the laser power measured at each frequency using a monitor photodiode.

The fluorescence signal in Fig.\ \ref{fig:laser_induced_fluorescence} shows an inhomogeneously broadened zero-phonon line, which we attribute to the $^{147,149}$Sm isotopes: the ratio of the heights of the two peak-like features matches the isotopic abundance ratio of $^{147}$Sm and $^{149}$Sm (1.09:1), and the width of these features is comparable to the inhomogeneous broadening of the \Clocktrans\ transition in $^{151}$Eu:YSO at similar doping concentration \cite{konz2003temperature}. However, the inhomogeneous profile is evidently non-gaussian, indicating that the statistical distribution of the interactions between the fluorescing ions and their local lattice environment is markedly different in Sm:\srf compared to Eu:YSO \cite{stoneham1969shapes}. We did not detect any phonon sidebands of the transition at frequencies up to 2 THz away from the zero-phonon line. This suggests that the \Clocktrans\ transition is only weakly coupled to the \ce{SrF_2} lattice.

\begin{figure}[h!]
   \includegraphics[width=\columnwidth]{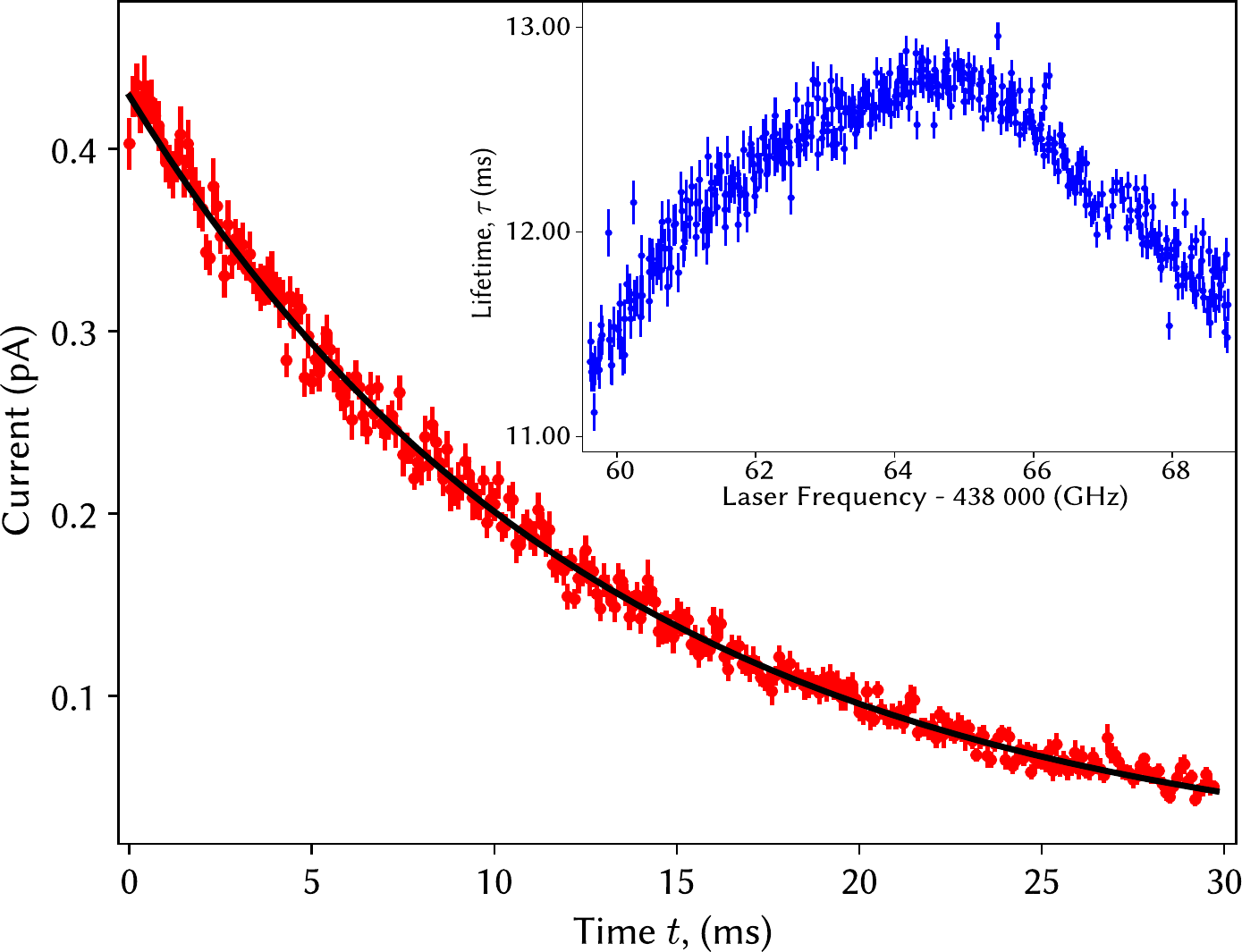}
  \caption{The decay of fluorescence at 697 nm observed after pulsing the probe laser (stabilized at 438065.5 GHz for this measurement). The solid line is a fit to a single exponential with a $1/e$ lifetime $\tau = 12.96(7)$ ms. The inset shows $\tau$ for different spectral classes within the inhomogeneous profile.}
\label{fig:RedDecay}
\end{figure}

The lifetime of the excited state was measured by fitting an exponential decay curve to the fluorescence signal measured in the dark. The decay fits very well to a single exponential, as shown in Fig.\ \ref{fig:RedDecay}. The inset to Fig.\ \ref{fig:RedDecay} shows the variation of the $^5D_0$ state lifetime $\tau$, as a function of the probe laser frequency. The lifetime varies by $\sim$10\% over the inhomogeneously broadened profile, potentially due to variation of the matrix element for the $^5D_0 \to ~^7F_1$ transition over the range of environments represented by these spectral classes. The longest lifetimes are observed near the peaks in Fig.\ \ref{fig:laser_induced_fluorescence}. (We also measured the lifetime of the $^5D_0$ state by indirectly populating it using a 410 nm laser, via the higher excited $4f^5 5d$ band, similar to the method used by Alam \emph{et al.} in Ref.\ \cite{mahbubul1967lifetime}. However, these decay curves did not fit well to single exponentials, likely due to excitation of many spectral classes from the inhomogeneous distribution in the excitation to the broad $4f^5 5d$ band -- such indirect measurements of the lifetime are therefore inaccurate.) The lifetime of the $^5D_0$ state measured by direct excitation of the \Clocktrans\ transition (averaged across the spectral classes between 438060-438068 GHz) is $\tau$ = 12.4(3) ms. The corresponding lower limit to the homogeneous linewidth of the transition is $\gamma_{h,\mrm{min}} \approx 2\pi \times 13$ Hz. 


\begin{figure}[h!]
\includegraphics[width = \columnwidth]{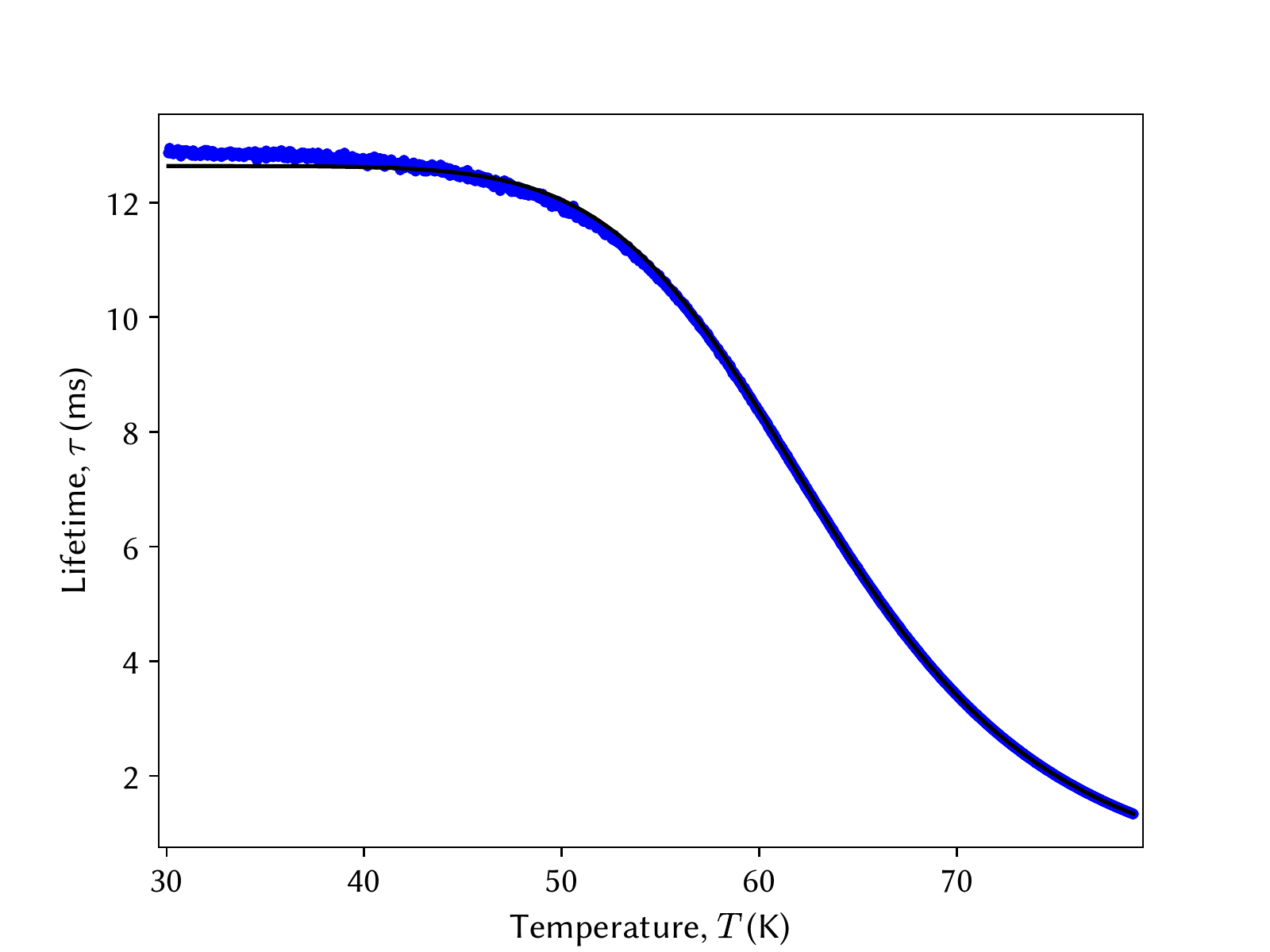}
\centering
\caption{The lifetime of the \e state as a function of temperature. The solid line is a fit to the thermally-activated decay model in Eq.\ (\ref{eq:thermally_activated_decay}).
} 
\label{fig:HeatUp}
\end{figure}

We also measured the lifetime of the \e state as a function of the temperature of the crystal, as shown in Fig.\ \ref{fig:HeatUp}. The lifetime fits well (albeit with a small discrepancy at low temperatures) to a simple model for a thermally-activated decay process \cite{mahbubul1967lifetime},
\begin{equation}\label{eq:thermally_activated_decay}
\frac{1}{\tau} = \frac{1}{\tau_1} + \frac{\exp(-\Delta/k_B T)}{\tau_2},
\end{equation}
where $\tau_1$ is the lifetime of the $^5D_0$ state at absolute zero, and $\tau_2$ is the lifetime of a shorter-lived state lying at an energy $\Delta$ above the $^5D_0$ state. The fit shown in Fig. \ref{fig:HeatUp} yields $\tau_2 = 200$ ns and $\Delta = h \times 14.2$ THz. This energy separation $\Delta$ is consistent with an experimentally observed state that is $\sim$450 cm$^{-1}$ above the $^5D_0$ state \cite{wood1962absorption}. 

\begin{figure}[h!]
\includegraphics[width = \columnwidth]{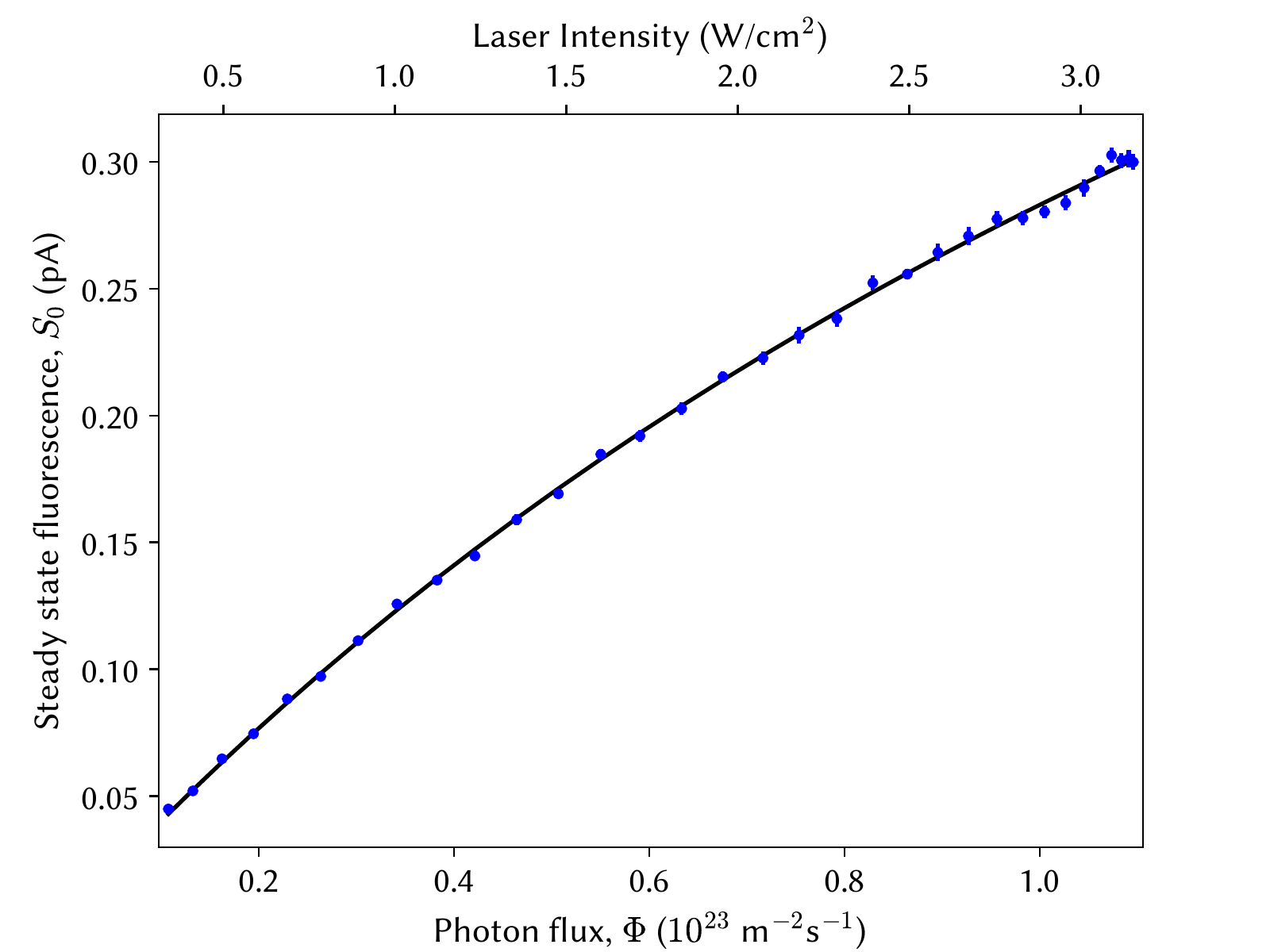}
\centering
\caption{The steady state fluorescence signal plotted against laser intensity. The error bars are the standard deviations of 16 repeated measurements of $S_0$ for each value of $I$. The solid line is a fit to Eq.(\ref{eq:crossSection}), which yields the excitation cross section.}
\label{fig:CrossSection}
\end{figure}

In contrast to the \Clocktrans\ transition in Eu:YSO at comparable REI densities (e.g., \cite{yano1991ultralong}), we did not observe any significant absorption of the probe laser through the crystal, indicating that the transition in Sm:\ce{SrF_2} is extremely weak. To quantify the line strength, we measured the excitation cross section for the \Clocktrans\ transition as follows. Using a frequency-stabilized laser at $\nu = 438066.1$ GHz, we pulsed the probe laser and measured both the rise and decay of the $^5D_0 \to {}^7F_1$ fluorescence as shown in Fig.\ \ref{fig:saturation}. The exponential decay yields the lifetime $\tau$ of the $^5D_0$ state as described above. The rising exponential was fit to the function $S(t) = S_0 \, (1 - e^{-t/\tau}) + b $ to extract the steady-state fluorescence signal $S_0$. We then repeated these measurements at a number of values of the laser intensity $I$ (using a beam with 100 $\mu$m 1/e$^2$ radius), to obtain $S_0(I)$ as shown in Fig.\ \ref{fig:CrossSection}. The nonlinearity in Fig.\ \ref{fig:CrossSection} indicates the onset of saturation on the \Clocktrans\ transition.

We modeled the intensity dependence of the steady-state fluorescence signal $S_0(I)$ using a set of rate equations for the three energy levels shown in Fig.\ \ref{fig:LevelStructure}(a),
\begin{equation}
\frac{d}{dt}
\begin{pmatrix}
N_g \\
N_e \\
N_f
\end{pmatrix} = 
\begin{pmatrix}
-(N_g - N_e) \Gamma_\mrm{exc} + \gamma_f N_f \\
(N_g - N_e)\Gamma_\mrm{exc} - \gamma_e N_e \\
\gamma_e N_e - \gamma_f N_f
\end{pmatrix}. 
\label{eq:rateeqs}
\end{equation}
Here $N_g, N_e, N_f$ are the populations in the $^7F_0, {}^5D_0, {}^7F_1$ states respectively. The excitation rate from $^7F_0 \to {}^5D_0$ is $\Gamma_\mrm{exc} = \Phi \sigma$, where $\Phi = I/h\nu$ is the photon flux and $\sigma$ is the excitation cross section of interest. The  decay rates for the $^5D_0 \to {}^7F_1$ and $^7F_1 \to {}^7F_0$ transitions are $\gamma_e$ and $\gamma_f$ respectively. $N = N_g + N_e + N_f$ is the total number of ions that are resonant with the laser. 

The measured steady-state fluorescence $S_0$ is related to the $^5D_0$ population at steady state, $N_{e,\mrm{ss}}$, via the decay rate $\gamma_e$, and an experimental efficiency factor $\eta_c$ (which accounts for the solid-angle subtended by the detector, the detector's efficiency, etc.). The resulting relation between $S_0$ and the cross section $\sigma$ is 
\begin{equation}\label{eq:crossSection}
\begin{split}
    S_0 & = \eta_c \gamma_e N_{e,\mrm{ss}} = \eta_c \gamma_e N \frac{\Phi \sigma}{\Phi \sigma \left(2 + \frac{\gamma_e}{\gamma_f} \right) +\gamma_e}. 
\end{split}
\end{equation}
We fix $\gamma_e = 1/\tau$ to its measured value, and assume that $\gamma_f \gg \gamma_e$, since the $^7F_1$ state is expected to relax quickly to the ground state due to the strong overlap between its energy above the ground state ($h \times$ 7.8 THz) and the phonon spectrum in \ce{SrF_2} \cite{kaiser1962infrared}. This assumption is borne out by the excellent fit of Eq.(\ref{eq:crossSection}) to the data in Fig.\ \ref{fig:CrossSection} (reduced $\chi^2$ = 0.98). The fit also yields a value for $\eta_c N$ which is consistent with an independent estimate of this quantity based on the calculated number of ions within the laser excitation volume that are resonant with the laser, manufacturer specifications for the filter transmissions and detector efficiency, and Monte Carlo modeling of the solid angle for light collection.
The resulting excitation cross section is $\sigma = 1.9(1) \times 10^{-18}$ cm$^2$, where the uncertainty is the quadrature sum of the fit uncertainty and the estimated intensity calibration uncertainty. This small cross section confirms the highly forbidden nature of the \Clocktrans\ transition in Sm:\ce{SrF_2}. 

\section{Discussion and summary}
It is interesting to ask why the transition between the $^7F_0$ and $^5D_0$ states is allowed at all in an octahedrally symmetric substitution site in \ce{SrF_2}, where it ought to be completely forbidden \cite{tanner2013some}. Based on the excitation cross section measured above, we suggest that the reason is hyperfine-induced mixing of the $^7F_0$ and $^7F_1$ states. In a magnetic field $\Bsca_\mrm{n} \sim 100$ G due to the Sm nucleus, the $^7F_0$ state is perturbed to $\ket{\widetilde{^7F_0}} \approx \ket{^7F_0} + \xi  \ket{^7F_1}$, where $\xi = \left(\mu \Bsca_\mrm{n}/{\Delta E} \right)$ with $\mu = 2 \mu_B$ being the $M1$ matrix element between the $^7F_0, {}^7F_1$ states and $\Delta E = h \times$ 7.8 THz their energy separation. This leads to a hyperfine-induced mixing amplitude $\xi \sim 4 \times 10^{-5}$, and an estimated homogeneous excitation cross section $\sigma_\mrm{hyp} = \xi^2 \, \frac{\lambda^2}{2\pi} \sim 10^{-18}$ cm$^2$ which is comparable to the experimentally measured cross section. 

The above estimate lends further credence to our suggestion that the inhomogeneously broadened lines shown in Fig.\ \ref{fig:laser_induced_fluorescence} arise from the $^{147,149}$Sm isotopes that have nonzero nuclear magnetic moments. Despite this inhomogeneous broadening, it may be possible to use the $^{147,149}$Sm lines for laser stabilization using spectral hole-burning similar to Eu:YSO, although burning persistent spectral holes in Sm:\ce{SrF_2} will require significantly higher optical intensities compared to Eu:YSO due to the relative weakness of the transition.

We expect that the zero spin isotopes of Sm will exhibit lower inhomogeneous broadening due to their reduced interactions with the lattice, although the clock transition in these isotopes will be extremely forbidden in the absence of hyperfine-induced mixing. The observation of the zero spin isotope lines is challenging for this reason, but is necessary in order to realize an absolute frequency reference using the Sm:\ce{SrF_2} system. The observation of these lines may be possible using isotopically enriched samples, or large magnetic fields to admix the $^7F_0,{}^7F_1$ states. We are exploring both these approaches. 


In summary, we have directly observed the highly forbidden \Clocktrans\ transition in Sm:\ce{SrF_2}. The population dynamics of the clock states under laser excitation have been understood quantitatively, allowing us to measure the lifetime of the excited state, and determine the mechanism that makes this transition slightly allowed. Our observation of this forbidden transition enables further studies of the properties of Sm:\ce{SrF_2}, to evaluate its feasibility as a radically simple optical frequency reference.

~ \\
\emph{Acknowledgments. --} We acknowledge the contributions of Graham Edge, Shreyas Potnis, Wesley Cassidy and Harish Ramachandran to this experiment. We acknowledge helpful discussions with Jonathan Weinstein and Andrew Jayich. This investigation was supported by funding from NSERC, CFI, Canada Research Chairs and the Branco Weiss Fellowship. M.V., L.T., and H.S. acknowledge support from NSERC Undergraduate Student Research Awards.

\bibliography{sm_observation}

\begin{thebibliography}{27}%
\makeatletter
\providecommand \@ifxundefined [1]{%
 \@ifx{#1\undefined}
}%
\providecommand \@ifnum [1]{%
 \ifnum #1\expandafter \@firstoftwo
 \else \expandafter \@secondoftwo
 \fi
}%
\providecommand \@ifx [1]{%
 \ifx #1\expandafter \@firstoftwo
 \else \expandafter \@secondoftwo
 \fi
}%
\providecommand \natexlab [1]{#1}%
\providecommand \enquote  [1]{``#1''}%
\providecommand \bibnamefont  [1]{#1}%
\providecommand \bibfnamefont [1]{#1}%
\providecommand \citenamefont [1]{#1}%
\providecommand \href@noop [0]{\@secondoftwo}%
\providecommand \href [0]{\begingroup \@sanitize@url \@href}%
\providecommand \@href[1]{\@@startlink{#1}\@@href}%
\providecommand \@@href[1]{\endgroup#1\@@endlink}%
\providecommand \@sanitize@url [0]{\catcode `\\12\catcode `\$12\catcode
  `\&12\catcode `\#12\catcode `\^12\catcode `\_12\catcode `\%12\relax}%
\providecommand \@@startlink[1]{}%
\providecommand \@@endlink[0]{}%
\providecommand \url  [0]{\begingroup\@sanitize@url \@url }%
\providecommand \@url [1]{\endgroup\@href {#1}{\urlprefix }}%
\providecommand \urlprefix  [0]{URL }%
\providecommand \Eprint [0]{\href }%
\providecommand \doibase [0]{http://dx.doi.org/}%
\providecommand \selectlanguage [0]{\@gobble}%
\providecommand \bibinfo  [0]{\@secondoftwo}%
\providecommand \bibfield  [0]{\@secondoftwo}%
\providecommand \translation [1]{[#1]}%
\providecommand \BibitemOpen [0]{}%
\providecommand \bibitemStop [0]{}%
\providecommand \bibitemNoStop [0]{.\EOS\space}%
\providecommand \EOS [0]{\spacefactor3000\relax}%
\providecommand \BibitemShut  [1]{\csname bibitem#1\endcsname}%
\let\auto@bib@innerbib\@empty
\bibitem [{\citenamefont {Rubio}(1991)}]{rubio1991doubly}%
  \BibitemOpen
  \bibfield  {author} {\bibinfo {author} {\bibfnamefont {O.}~\bibnamefont
  {Rubio}},\ }\href@noop {} {\bibfield  {journal} {\bibinfo  {journal} {J.
  Phys. Chem. Sol.}\ }\textbf {\bibinfo {volume} {52}},\ \bibinfo {pages} {101}
  (\bibinfo {year} {1991})}\BibitemShut {NoStop}%
\bibitem [{\citenamefont {Wybourne}(2004)}]{Wybourne2004}%
  \BibitemOpen
  \bibfield  {author} {\bibinfo {author} {\bibfnamefont {B.~G.}\ \bibnamefont
  {Wybourne}},\ }\href@noop {} {\bibfield  {journal} {\bibinfo  {journal} {J.
  Alloys Compd.}\ }\textbf {\bibinfo {volume} {380}},\ \bibinfo {pages} {96}
  (\bibinfo {year} {2004})}\BibitemShut {NoStop}%
\bibitem [{\citenamefont {Hull}\ \emph {et~al.}(2005)\citenamefont {Hull},
  \citenamefont {Parisi}, \citenamefont {Osgood}, \citenamefont {Warlimont},
  \citenamefont {Liu},\ and\ \citenamefont {Jacquier}}]{Hull2005}%
  \BibitemOpen
  \bibinfo {editor} {\bibfnamefont {R.}~\bibnamefont {Hull}}, \bibinfo {editor}
  {\bibfnamefont {J.}~\bibnamefont {Parisi}}, \bibinfo {editor} {\bibfnamefont
  {R.~M.}\ \bibnamefont {Osgood}}, \bibinfo {editor} {\bibfnamefont
  {H.}~\bibnamefont {Warlimont}}, \bibinfo {editor} {\bibfnamefont
  {G.}~\bibnamefont {Liu}}, \ and\ \bibinfo {editor} {\bibfnamefont
  {B.}~\bibnamefont {Jacquier}},\ eds.,\ \href {\doibase 10.1007/3-540-28209-2}
  {\emph {\bibinfo {title} {{Spectroscopic Properties of Rare Earths in Optical
  Materials}}}},\ \bibinfo {series} {Springer Series in Materials Science},
  Vol.~\bibinfo {volume} {83}\ (\bibinfo  {publisher} {Springer-Verlag},\
  \bibinfo {address} {Berlin/Heidelberg},\ \bibinfo {year} {2005})\BibitemShut
  {NoStop}%
\bibitem [{\citenamefont {Simon}\ \emph {et~al.}(2010)\citenamefont {Simon}
  \emph {et~al.}}]{simon2010quantum}%
  \BibitemOpen
  \bibfield  {author} {\bibinfo {author} {\bibfnamefont {C.}~\bibnamefont
  {Simon}} \emph {et~al.},\ }\href@noop {} {\bibfield  {journal} {\bibinfo
  {journal} {Eur. Phys. J. D}\ }\textbf {\bibinfo {volume} {58}},\ \bibinfo
  {pages} {1} (\bibinfo {year} {2010})}\BibitemShut {NoStop}%
\bibitem [{\citenamefont {Thiel}\ \emph {et~al.}(2011)\citenamefont {Thiel},
  \citenamefont {B{\"o}ttger},\ and\ \citenamefont {Cone}}]{thiel2011rare}%
  \BibitemOpen
  \bibfield  {author} {\bibinfo {author} {\bibfnamefont {C.}~\bibnamefont
  {Thiel}}, \bibinfo {author} {\bibfnamefont {T.}~\bibnamefont {B{\"o}ttger}},
  \ and\ \bibinfo {author} {\bibfnamefont {R.}~\bibnamefont {Cone}},\
  }\href@noop {} {\bibfield  {journal} {\bibinfo  {journal} {J. Lumin.}\
  }\textbf {\bibinfo {volume} {131}},\ \bibinfo {pages} {353} (\bibinfo {year}
  {2011})}\BibitemShut {NoStop}%
\bibitem [{\citenamefont {Zhong}\ \emph {et~al.}(2015)\citenamefont {Zhong},
  \citenamefont {Hedges}, \citenamefont {Ahlefeldt}, \citenamefont
  {Bartholomew}, \citenamefont {Beavan}, \citenamefont {Wittig}, \citenamefont
  {Longdell},\ and\ \citenamefont {Sellars}}]{Zhong2015}%
  \BibitemOpen
  \bibfield  {author} {\bibinfo {author} {\bibfnamefont {M.}~\bibnamefont
  {Zhong}}, \bibinfo {author} {\bibfnamefont {M.~P.}\ \bibnamefont {Hedges}},
  \bibinfo {author} {\bibfnamefont {R.~L.}\ \bibnamefont {Ahlefeldt}}, \bibinfo
  {author} {\bibfnamefont {J.~G.}\ \bibnamefont {Bartholomew}}, \bibinfo
  {author} {\bibfnamefont {S.~E.}\ \bibnamefont {Beavan}}, \bibinfo {author}
  {\bibfnamefont {S.~M.}\ \bibnamefont {Wittig}}, \bibinfo {author}
  {\bibfnamefont {J.~J.}\ \bibnamefont {Longdell}}, \ and\ \bibinfo {author}
  {\bibfnamefont {M.~J.}\ \bibnamefont {Sellars}},\ }\href@noop {} {\bibfield
  {journal} {\bibinfo  {journal} {Nature}\ }\textbf {\bibinfo {volume} {517}},\
  \bibinfo {pages} {177} (\bibinfo {year} {2015})}\BibitemShut {NoStop}%
\bibitem [{\citenamefont {Dibos}\ \emph {et~al.}(2018)\citenamefont {Dibos},
  \citenamefont {Raha}, \citenamefont {Phenicie},\ and\ \citenamefont
  {Thompson}}]{dibos2018atomic}%
  \BibitemOpen
  \bibfield  {author} {\bibinfo {author} {\bibfnamefont {A.}~\bibnamefont
  {Dibos}}, \bibinfo {author} {\bibfnamefont {M.}~\bibnamefont {Raha}},
  \bibinfo {author} {\bibfnamefont {C.}~\bibnamefont {Phenicie}}, \ and\
  \bibinfo {author} {\bibfnamefont {J.}~\bibnamefont {Thompson}},\ }\href@noop
  {} {\bibfield  {journal} {\bibinfo  {journal} {Phys. Rev. Lett.}\ }\textbf
  {\bibinfo {volume} {120}},\ \bibinfo {pages} {243601} (\bibinfo {year}
  {2018})}\BibitemShut {NoStop}%
\bibitem [{\citenamefont {Nilsson}\ and\ \citenamefont
  {Kr{\"o}ll}(2005)}]{nilsson2005solid}%
  \BibitemOpen
  \bibfield  {author} {\bibinfo {author} {\bibfnamefont {M.}~\bibnamefont
  {Nilsson}}\ and\ \bibinfo {author} {\bibfnamefont {S.}~\bibnamefont
  {Kr{\"o}ll}},\ }\href@noop {} {\bibfield  {journal} {\bibinfo  {journal}
  {Opt. Commun.}\ }\textbf {\bibinfo {volume} {247}},\ \bibinfo {pages} {393}
  (\bibinfo {year} {2005})}\BibitemShut {NoStop}%
\bibitem [{\citenamefont {Thorpe}\ \emph {et~al.}(2011)\citenamefont {Thorpe},
  \citenamefont {Rippe}, \citenamefont {Fortier}, \citenamefont {Kirchner},\
  and\ \citenamefont {Rosenband}}]{thorpe2011frequency}%
  \BibitemOpen
  \bibfield  {author} {\bibinfo {author} {\bibfnamefont {M.~J.}\ \bibnamefont
  {Thorpe}}, \bibinfo {author} {\bibfnamefont {L.}~\bibnamefont {Rippe}},
  \bibinfo {author} {\bibfnamefont {T.~M.}\ \bibnamefont {Fortier}}, \bibinfo
  {author} {\bibfnamefont {M.~S.}\ \bibnamefont {Kirchner}}, \ and\ \bibinfo
  {author} {\bibfnamefont {T.}~\bibnamefont {Rosenband}},\ }\href@noop {}
  {\bibfield  {journal} {\bibinfo  {journal} {Nat. Photonics}\ }\textbf
  {\bibinfo {volume} {5}},\ \bibinfo {pages} {688} (\bibinfo {year}
  {2011})}\BibitemShut {NoStop}%
\bibitem [{\citenamefont {Cook}\ \emph {et~al.}(2015)\citenamefont {Cook},
  \citenamefont {Rosenband},\ and\ \citenamefont {Leibrandt}}]{Cook2015}%
  \BibitemOpen
  \bibfield  {author} {\bibinfo {author} {\bibfnamefont {S.}~\bibnamefont
  {Cook}}, \bibinfo {author} {\bibfnamefont {T.}~\bibnamefont {Rosenband}}, \
  and\ \bibinfo {author} {\bibfnamefont {D.~R.}\ \bibnamefont {Leibrandt}},\
  }\href@noop {} {\bibfield  {journal} {\bibinfo  {journal} {Phys. Rev. Lett.}\
  }\textbf {\bibinfo {volume} {114}},\ \bibinfo {pages} {253902} (\bibinfo
  {year} {2015})}\BibitemShut {NoStop}%
\bibitem [{\citenamefont {Equall}\ \emph {et~al.}(1994)\citenamefont {Equall},
  \citenamefont {Sun}, \citenamefont {Cone},\ and\ \citenamefont
  {Macfarlane}}]{equall1994ultraslow}%
  \BibitemOpen
  \bibfield  {author} {\bibinfo {author} {\bibfnamefont {R.~W.}\ \bibnamefont
  {Equall}}, \bibinfo {author} {\bibfnamefont {Y.}~\bibnamefont {Sun}},
  \bibinfo {author} {\bibfnamefont {R.}~\bibnamefont {Cone}}, \ and\ \bibinfo
  {author} {\bibfnamefont {R.}~\bibnamefont {Macfarlane}},\ }\href@noop {}
  {\bibfield  {journal} {\bibinfo  {journal} {Phys. Rev. Lett.}\ }\textbf
  {\bibinfo {volume} {72}},\ \bibinfo {pages} {2179} (\bibinfo {year}
  {1994})}\BibitemShut {NoStop}%
\bibitem [{\citenamefont {Siyushev}\ \emph {et~al.}(2014)\citenamefont
  {Siyushev} \emph {et~al.}}]{siyushev2014coherent}%
  \BibitemOpen
  \bibfield  {author} {\bibinfo {author} {\bibfnamefont {P.}~\bibnamefont
  {Siyushev}} \emph {et~al.},\ }\href@noop {} {\bibfield  {journal} {\bibinfo
  {journal} {Nat. Commun.}\ }\textbf {\bibinfo {volume} {5}},\ \bibinfo {pages}
  {3895} (\bibinfo {year} {2014})}\BibitemShut {NoStop}%
\bibitem [{\citenamefont {K{\"o}nz}\ \emph {et~al.}(2003)\citenamefont
  {K{\"o}nz}, \citenamefont {Sun}, \citenamefont {Thiel}, \citenamefont {Cone},
  \citenamefont {Equall}, \citenamefont {Hutcheson},\ and\ \citenamefont
  {Macfarlane}}]{konz2003temperature}%
  \BibitemOpen
  \bibfield  {author} {\bibinfo {author} {\bibfnamefont {F.}~\bibnamefont
  {K{\"o}nz}}, \bibinfo {author} {\bibfnamefont {Y.}~\bibnamefont {Sun}},
  \bibinfo {author} {\bibfnamefont {C.}~\bibnamefont {Thiel}}, \bibinfo
  {author} {\bibfnamefont {R.}~\bibnamefont {Cone}}, \bibinfo {author}
  {\bibfnamefont {R.}~\bibnamefont {Equall}}, \bibinfo {author} {\bibfnamefont
  {R.}~\bibnamefont {Hutcheson}}, \ and\ \bibinfo {author} {\bibfnamefont
  {R.}~\bibnamefont {Macfarlane}},\ }\href@noop {} {\bibfield  {journal}
  {\bibinfo  {journal} {Phys. Rev. B}\ }\textbf {\bibinfo {volume} {68}},\
  \bibinfo {pages} {085109} (\bibinfo {year} {2003})}\BibitemShut {NoStop}%
\bibitem [{\citenamefont {Leibrandt}\ \emph {et~al.}(2013)\citenamefont
  {Leibrandt}, \citenamefont {Thorpe}, \citenamefont {Chou}, \citenamefont
  {Fortier}, \citenamefont {Diddams},\ and\ \citenamefont
  {Rosenband}}]{leibrandt2013absolute}%
  \BibitemOpen
  \bibfield  {author} {\bibinfo {author} {\bibfnamefont {D.~R.}\ \bibnamefont
  {Leibrandt}}, \bibinfo {author} {\bibfnamefont {M.~J.}\ \bibnamefont
  {Thorpe}}, \bibinfo {author} {\bibfnamefont {C.-W.}\ \bibnamefont {Chou}},
  \bibinfo {author} {\bibfnamefont {T.~M.}\ \bibnamefont {Fortier}}, \bibinfo
  {author} {\bibfnamefont {S.~A.}\ \bibnamefont {Diddams}}, \ and\ \bibinfo
  {author} {\bibfnamefont {T.}~\bibnamefont {Rosenband}},\ }\href@noop {}
  {\bibfield  {journal} {\bibinfo  {journal} {Phys. Rev. Lett.}\ }\textbf
  {\bibinfo {volume} {111}},\ \bibinfo {pages} {237402} (\bibinfo {year}
  {2013})}\BibitemShut {NoStop}%
\bibitem [{\citenamefont {Ferrier}\ \emph {et~al.}(2016)\citenamefont
  {Ferrier}, \citenamefont {Tumino},\ and\ \citenamefont
  {Goldner}}]{Ferrier2016}%
  \BibitemOpen
  \bibfield  {author} {\bibinfo {author} {\bibfnamefont {A.}~\bibnamefont
  {Ferrier}}, \bibinfo {author} {\bibfnamefont {B.}~\bibnamefont {Tumino}}, \
  and\ \bibinfo {author} {\bibfnamefont {P.}~\bibnamefont {Goldner}},\
  }\href@noop {} {\bibfield  {journal} {\bibinfo  {journal} {J. Lumin.}\
  }\textbf {\bibinfo {volume} {170}},\ \bibinfo {pages} {406} (\bibinfo {year}
  {2016})}\BibitemShut {NoStop}%
\bibitem [{\citenamefont {Stoneham}(1969)}]{stoneham1969shapes}%
  \BibitemOpen
  \bibfield  {author} {\bibinfo {author} {\bibfnamefont {A.}~\bibnamefont
  {Stoneham}},\ }\href@noop {} {\bibfield  {journal} {\bibinfo  {journal} {Rev.
  Mod. Phys.}\ }\textbf {\bibinfo {volume} {41}},\ \bibinfo {pages} {82}
  (\bibinfo {year} {1969})}\BibitemShut {NoStop}%
\bibitem [{\citenamefont {Oswald}\ \emph {et~al.}(2018)\citenamefont {Oswald},
  \citenamefont {Hansen}, \citenamefont {Wiens}, \citenamefont {Nevsky},\ and\
  \citenamefont {Schiller}}]{oswald2018characteristics}%
  \BibitemOpen
  \bibfield  {author} {\bibinfo {author} {\bibfnamefont {R.}~\bibnamefont
  {Oswald}}, \bibinfo {author} {\bibfnamefont {M.~G.}\ \bibnamefont {Hansen}},
  \bibinfo {author} {\bibfnamefont {E.}~\bibnamefont {Wiens}}, \bibinfo
  {author} {\bibfnamefont {A.~Y.}\ \bibnamefont {Nevsky}}, \ and\ \bibinfo
  {author} {\bibfnamefont {S.}~\bibnamefont {Schiller}},\ }\href@noop {}
  {\bibfield  {journal} {\bibinfo  {journal} {Phys. Rev. A}\ }\textbf {\bibinfo
  {volume} {98}},\ \bibinfo {pages} {062516} (\bibinfo {year}
  {2018})}\BibitemShut {NoStop}%
\bibitem [{\citenamefont {Wasserburg}\ \emph {et~al.}(1981)\citenamefont
  {Wasserburg}, \citenamefont {Jacobsen}, \citenamefont {DePaolo},
  \citenamefont {McCulloch},\ and\ \citenamefont {Wen}}]{Wasserburg1981}%
  \BibitemOpen
  \bibfield  {author} {\bibinfo {author} {\bibfnamefont {G.}~\bibnamefont
  {Wasserburg}}, \bibinfo {author} {\bibfnamefont {S.}~\bibnamefont
  {Jacobsen}}, \bibinfo {author} {\bibfnamefont {D.}~\bibnamefont {DePaolo}},
  \bibinfo {author} {\bibfnamefont {M.}~\bibnamefont {McCulloch}}, \ and\
  \bibinfo {author} {\bibfnamefont {T.}~\bibnamefont {Wen}},\ }\href {\doibase
  10.1016/0016-7037(81)90085-5} {\bibfield  {journal} {\bibinfo  {journal}
  {Geochimica et Cosmochimica Acta}\ }\textbf {\bibinfo {volume} {45}},\
  \bibinfo {pages} {2311} (\bibinfo {year} {1981})}\BibitemShut {NoStop}%
\bibitem [{\citenamefont {Jia}(1991)}]{Jia1991}%
  \BibitemOpen
  \bibfield  {author} {\bibinfo {author} {\bibfnamefont {Y.}~\bibnamefont
  {Jia}},\ }\href@noop {} {\bibfield  {journal} {\bibinfo  {journal} {J. Solid
  State Chem.}\ }\textbf {\bibinfo {volume} {95}},\ \bibinfo {pages} {184 }
  (\bibinfo {year} {1991})}\BibitemShut {NoStop}%
\bibitem [{\citenamefont {Shannon}(1976)}]{Shannon1976}%
  \BibitemOpen
  \bibfield  {author} {\bibinfo {author} {\bibfnamefont {R.~D.}\ \bibnamefont
  {Shannon}},\ }\href {\doibase 10.1107/S0567739476001551} {\bibfield
  {journal} {\bibinfo  {journal} {Acta Crystallographica A}\ }\textbf {\bibinfo
  {volume} {32}},\ \bibinfo {pages} {751} (\bibinfo {year} {1976})}\BibitemShut
  {NoStop}%
\bibitem [{\citenamefont {Wood}\ and\ \citenamefont
  {Kaiser}(1962)}]{wood1962absorption}%
  \BibitemOpen
  \bibfield  {author} {\bibinfo {author} {\bibfnamefont {D.}~\bibnamefont
  {Wood}}\ and\ \bibinfo {author} {\bibfnamefont {W.}~\bibnamefont {Kaiser}},\
  }\href@noop {} {\bibfield  {journal} {\bibinfo  {journal} {Phys. Rev.}\
  }\textbf {\bibinfo {volume} {126}},\ \bibinfo {pages} {2079} (\bibinfo {year}
  {1962})}\BibitemShut {NoStop}%
\bibitem [{\citenamefont {Alam}\ and\ \citenamefont
  {Di~Bartolo}(1967)}]{mahbubul1967lifetime}%
  \BibitemOpen
  \bibfield  {author} {\bibinfo {author} {\bibfnamefont {A.~M.}\ \bibnamefont
  {Alam}}\ and\ \bibinfo {author} {\bibfnamefont {B.}~\bibnamefont
  {Di~Bartolo}},\ }\href@noop {} {\bibfield  {journal} {\bibinfo  {journal}
  {Phys. Lett. A}\ }\textbf {\bibinfo {volume} {25}},\ \bibinfo {pages} {157}
  (\bibinfo {year} {1967})}\BibitemShut {NoStop}%
\bibitem [{\citenamefont {G{\^a}con}\ \emph {et~al.}(1993)\citenamefont
  {G{\^a}con}, \citenamefont {Burdick}, \citenamefont {Moine},\ and\
  \citenamefont {Bill}}]{gacon1993f}%
  \BibitemOpen
  \bibfield  {author} {\bibinfo {author} {\bibfnamefont {J.-C.}\ \bibnamefont
  {G{\^a}con}}, \bibinfo {author} {\bibfnamefont {G.}~\bibnamefont {Burdick}},
  \bibinfo {author} {\bibfnamefont {B.}~\bibnamefont {Moine}}, \ and\ \bibinfo
  {author} {\bibfnamefont {H.}~\bibnamefont {Bill}},\ }\href@noop {} {\bibfield
   {journal} {\bibinfo  {journal} {Phys. Rev. B}\ }\textbf {\bibinfo {volume}
  {47}},\ \bibinfo {pages} {11712} (\bibinfo {year} {1993})}\BibitemShut
  {NoStop}%
\bibitem [{\citenamefont {Macfarlane}\ and\ \citenamefont
  {Meltzer}(1985)}]{macfarlane1985spectral}%
  \BibitemOpen
  \bibfield  {author} {\bibinfo {author} {\bibfnamefont {R.}~\bibnamefont
  {Macfarlane}}\ and\ \bibinfo {author} {\bibfnamefont {R.}~\bibnamefont
  {Meltzer}},\ }\href@noop {} {\bibfield  {journal} {\bibinfo  {journal} {Opt.
  Commun.}\ }\textbf {\bibinfo {volume} {52}},\ \bibinfo {pages} {320}
  (\bibinfo {year} {1985})}\BibitemShut {NoStop}%
\bibitem [{\citenamefont {Tanner}(2013)}]{tanner2013some}%
  \BibitemOpen
  \bibfield  {author} {\bibinfo {author} {\bibfnamefont {P.~A.}\ \bibnamefont
  {Tanner}},\ }\href@noop {} {\bibfield  {journal} {\bibinfo  {journal} {Chem.
  Soc. Rev.}\ }\textbf {\bibinfo {volume} {42}},\ \bibinfo {pages} {5090}
  (\bibinfo {year} {2013})}\BibitemShut {NoStop}%
\bibitem [{\citenamefont {Yano}\ \emph {et~al.}(1991)\citenamefont {Yano},
  \citenamefont {Mitsunaga},\ and\ \citenamefont {Uesugi}}]{yano1991ultralong}%
  \BibitemOpen
  \bibfield  {author} {\bibinfo {author} {\bibfnamefont {R.}~\bibnamefont
  {Yano}}, \bibinfo {author} {\bibfnamefont {M.}~\bibnamefont {Mitsunaga}}, \
  and\ \bibinfo {author} {\bibfnamefont {N.}~\bibnamefont {Uesugi}},\
  }\href@noop {} {\bibfield  {journal} {\bibinfo  {journal} {Opt. Lett.}\
  }\textbf {\bibinfo {volume} {16}},\ \bibinfo {pages} {1884} (\bibinfo {year}
  {1991})}\BibitemShut {NoStop}%
\bibitem [{\citenamefont {Kaiser}\ \emph {et~al.}(1962)\citenamefont {Kaiser},
  \citenamefont {Spitzer}, \citenamefont {Kaiser},\ and\ \citenamefont
  {Howarth}}]{kaiser1962infrared}%
  \BibitemOpen
  \bibfield  {author} {\bibinfo {author} {\bibfnamefont {W.}~\bibnamefont
  {Kaiser}}, \bibinfo {author} {\bibfnamefont {W.}~\bibnamefont {Spitzer}},
  \bibinfo {author} {\bibfnamefont {R.}~\bibnamefont {Kaiser}}, \ and\ \bibinfo
  {author} {\bibfnamefont {L.}~\bibnamefont {Howarth}},\ }\href@noop {}
  {\bibfield  {journal} {\bibinfo  {journal} {Phys. Rev.}\ }\textbf {\bibinfo
  {volume} {127}},\ \bibinfo {pages} {1950} (\bibinfo {year}
  {1962})}\BibitemShut {NoStop}%
\end{thebibliography}%

\end{document}